\documentclass[12pt]{iopart}

\usepackage[colorlinks=true,urlcolor=blue,citecolor=blue,linkcolor=blue]{hyperref}

\usepackage{graphicx}
\usepackage{color}
\usepackage{changes}

\makeatletter
\renewcommand\@appendixstar{\@@par
 \ifnumbysec 
 \@addtoreset{table}{section}
 \@addtoreset{figure}{section}\fi
 \setcounter{section}{0}
 \setcounter{subsection}{0}
 \setcounter{subsubsection}{0}
 \setcounter{equation}{0}
 \setcounter{figure}{0}
 \setcounter{table}{0}
 \def\thesection{\Alph{section}} 
 \def\theequation{\ifnumbysec
      \Alph{section}.\arabic{equation}\else
      \Alph{section}\arabic{equation}\fi}
 \def\thetable{\ifnumbysec
      \Alph{section}\arabic{table}\else
      A\arabic{table}\fi}
 \def\thefigure{\ifnumbysec
      \Alph{section}\arabic{figure}\else
      A\arabic{figure}\fi}}
\makeatother

\begin{document}

\title[Optimizing qubit resources for quantum chemistry simulations]{Optimizing qubit resources for quantum chemistry simulations in second quantization on a quantum computer}
\author{Nikolaj Moll, Andreas Fuhrer, Peter Staar, Ivano Tavernelli}
\address{IBM Research -- Zurich, S\"aumerstrasse 4, 8803 R\"uschlikon, Switzerland}
\ead{nim@zurich.ibm.com,afu@zurich.ibm.com,taa@zurich.ibm.com, and ita@zurich.ibm.com}

\date{\today}

\begin{abstract}
Quantum chemistry simulations on a quantum computer suffer from the overhead needed for encoding the fermionic problem in a system of qubits. By exploiting the block diagonality of a fermionic Hamiltonian, we show that the number of required qubits can be reduced while the number of terms in the Hamiltonian will increase. All operations for this reduction can be performed in operator space. The scheme is conceived as a pre-computational step that would be performed prior to the actual quantum simulation. We apply this scheme to reduce the number of qubits necessary to simulate both the Hamiltonian of the two-site Fermi-Hubbard model and the hydrogen molecule. Both quantum systems can then be simulated with a two-qubit quantum computer. Despite the increase in the number of Hamiltonian terms, the scheme still remains a useful tool to reduce the dimensionality of specific quantum systems for quantum simulators with a limited number of resources.
\end{abstract}


\maketitle

\tableofcontents

\title[]{}

\section{Introduction}

Recent advances in the field of quantum computing have boosted the hope that one day we might be able to solve complex material-science problems using quantum computers \cite{abrams_quantum_1999, aspuru-guzik_simulated_2005, brown_using_2010, lanyon_towards_2010, kassal_simulating_2011, li_solving_2011, whitfield_simulation_2011, aspuru-guzik_photonic_2012, jones_faster_2012, seeley_bravyi-kitaev_2012, tempel_quantum_2012, whitfield_computational_2012, yung_quantumquantum_2012, toloui_quantum_2013, babbush_adiabatic_2014, peruzzo_variational_2014, poulin_trotter_2014, wecker_gate-count_2014, babbush_exponentially_2015, babbush_exponentially_2015-1, barends_digital_2015, garcia-alvarez_quantum_2015, hastings_improving_2015, omalley_scalable_2015, tranter_bravyikitaev_2015, whitfield_unified_2015}. It was shown that the direct mapping of the molecular wave function to the qubit state allows the unitary operator to be decomposed into a number of gates that only scales polynomially with system size~\cite{aspuru-guzik_simulated_2005}. A quantum computer could then be used for the simulation of chemical systems and their properties, including correlation functions and reaction rates~\cite{kassal_simulating_2011}. For example, the hydrogen molecule in a minimal basis was calculated with a photonic quantum computer \cite{lanyon_towards_2010, aspuru-guzik_photonic_2012} and a variational eigenvalue solver was demonstrated~\cite{peruzzo_variational_2014}. Beyond this, it was suggested that a small quantum computer with on the order of 100 qubits will be able to address challenging problems in quantum chemistry that are beyond the reach of classical algorithms~\cite{wecker_gate-count_2014}. Improvement of quantum algorithms~\cite{hastings_improving_2015}, such as the reduction of the number of trotter steps required, might facilitate this. Another path could be the direct application of the quantum adiabatic algorithm to the quantum computation of molecular properties~\cite{babbush_adiabatic_2014}.

The physics that govern the electrons in a material can be described by a many-body Hamiltonian written in its second quantization form,
\begin{equation}
H = \sum_{ij} h_{ij} c^\dagger_{i} c^{\phantom\dagger}_{j}+ \sum_{ijkl} h_{ijkl} c^{\dagger}_{i}c^{\dagger}_{j} c^{\phantom\dagger}_{k} c^{\phantom\dagger}_{l},
\label{eq:Hamiltonian}
\end{equation}
where the symbols $c^\dagger_{i}$ ($c_{i}$) represent fermionic creation (annihilation) operators of an electron in the atomic orbital $i$. The number of atomic orbitals $K$ sets the electronic degrees of freedom, and the size of the associated Hilbert space will be $2^{K}$ and the matrix size $2^{K} \times 2^{K}$. The coefficients $h_{ij}$ and $h_{ijkl}$ are the one-body interaction term and the two-body electron-electron interaction term, respectively. They result from overlap integrals that can be precomputed classically~\cite{whitfield_simulation_2011}. 

For an efficient quantum simulation of quantum chemistry or other fermionic systems on a quantum computer, first the original Hamiltonian has to be transformed because electrons are fermions~\cite{ortiz_quantum_2001}. The Jordan--Wigner~\cite{jordan_uber_1928} and the Bravyi--Kitaev~\cite{bravyi_fermionic_2002,seeley_bravyi-kitaev_2012} transformations are currently the most commonly used ones in the context of electronic-structure Hamiltonians. Here, we employ the Jordan--Wigner transform,
\begin{equation}
c^{\phantom\dagger}_{j} = \mathop{\Pi}\limits_{i=1}^{j-1} Z_i \left( X_j + i Y_j \right)
\,\, {\rm and} \,\,\, 
c^{\dagger}_{j} = \mathop{\Pi}\limits_{i=1}^{j-1} Z_i \left( X_j - i Y_j \right) \, ,
\end{equation}
where $Z_i$ is short-hand for the $\sigma_z$ Pauli operator of the qubit $i$ and multiplications are tensor products. The many-body interactions are increased from order 4 to $K$ for the Jordan--Wigner and to $\log_2 K$ for the Bravyi--Kitaev transformation. However, the size of the Hilbert space of the Hamiltonian remains $2^K$. It is therefore highly desirable to develop a new scheme that allows a reduction of the size of the relevant Hilbert space. 

Here, we derive such a scheme that reduces the dimensionality of the Hilbert space. A sensible approach is to restrict the Hamiltonian operator to the desired number of electrons $N$ as there is no interaction between the blocks corresponding to different numbers of electrons. The electronic Hamiltonian in second quantization has block diagonal form where each block corresponds to a total electron number $N$ between $0$ and $K$. By restricting the Hilbert space to one block, its dimensionality can be reduced from $2^{K}$ to $K \choose N$. For example, for a system with 10 electrons in 100 orbitals, the Hilbert space can be reduced to 44 qubits. A further reduction can be obtained when enforcing the total spin state of interest. In this case the Hamiltonian operator is restricted to the spaces span by the desired number of spin-up, $N_\uparrow$ and spin-down electrons $N_\downarrow$, separately. The size of the Hilbert space is therefore reduced from $2^K$ to ${K/2 \choose N_\uparrow} + {K/2 \choose N_\downarrow}$. In the case of the hydrogen molecule, we then apply the reduction to a single spin state (the singlet). Importantly, in our scheme there is no need to introduce a matrix representation for the Hamiltonian operator: all operations are performed in operator space. The scheme is a pre-computational step that would be performed prior the actual quantum simulation.

This paper is organized in the following way: first, we describe our scheme to reduce the dimension of the Hilbert space associated with the Hamiltonian in Eq.~(\ref{eq:Hamiltonian}) using the projector technique and the necessary qubit reduction operations without going into the basis. Then we will apply the scheme to two examples: the Fermi--Hubbard model with two sites and the hydrogen molecule with two orbitals.

\section{The General Scheme}

To restrict the Hamiltonian in Eq.~(\ref{eq:Hamiltonian}) to a fixed number of electrons $N$, we first project out all other blocks that do not correspond to the desired number of electrons by means of the operator
\begin{equation} 
P_{N}^{(K)} =  \prod_{j \neq N}^{K} P^{(K)}_{N,j} = \prod_{j \neq N}^{K} \frac{N_{\rm tot} - j}{N - j}
\label{eq:PNK}
\end{equation}
and
\begin{equation} \label{eq:HNK}
H_{N}^{(K)} = P_{N}^{{(K)} \dagger} H^{(K)} P_{N}^{(K) \phantom\dagger } \, .
\end{equation} 
The total number of electrons $N_{\rm tot}$ is the sum of occupations $n_j$ of all states
\begin{equation}
 N_{\rm tot} = \sum_{j = 1}^{K} n_j = \sum_{j = 1}^{K} c^\dagger_{j} c^{\phantom\dagger}_{j} \, .
\end{equation}
The Hilbert space of Hamiltonian $H_{N}^{(K)}$ is not yet reduced, but all blocks other than the block of interest are set to zero. This projection will lead to an increase of terms, which in the worst case, scales exponentially with $2^K$. Originally, the projector scales with $K^K$, however, the scaling can be reduced to $2^K$ by exploiting the idempotency $n_i^2 = n_i$ for occupation operator of each single state. In the following two examples we show that the number of terms is clearly smaller due to the sparsity of the Hamiltonian which is system dependent. Generally to keep the number of terms small, it may be beneficial not to evaluate the projector $P_{N}^{(K)}$ from Eq.~(\ref{eq:PNK}) in one piece but to evaluate it iteratively starting from the innermost factors
\begin{equation}
H_{N}^{(K)} = P^{(K)\dagger}_{N,K} \dots \left( P^{(K)\dagger}_{N,2} \left( P^{(K)\dagger}_{N,1} H^{(K)} P_{N,1}^{(K)} \right) P^{(K)}_{N,2} \right) \dots P^{(K)}_{N,K}
\end{equation}
and applying at each step the (anti)-commutation rules associated to the creation and annihilation operators.

The blocks set to zero can now be sequentially removed and the number of qubits reduced one by one. Exploiting the relationship that the tensor product of the reduced Hamiltonian $H_N^{(K-1)}$ (of dimension $2^{K-1})$ and the identity operator $1$ is equal to the direct sum of two copies of the reduced Hamiltonian $H_N^{(K-1)}$,
\begin{equation}
1 \otimes H_N^{(K-1)} = H_N^{(K-1)} \oplus H_N^{(K-1)} \, ,
\end{equation}
we can finally isolate the reduced Hamiltonian $H_N^{(K-1)}$. This operation is accomplished through the action of the shift operators $S_+$ and $S_-$ (see Appendix \ref{appendix:b})
\begin{equation} \label{eq:SpS-}
S_\pm^{(K)} = \frac{1}{2} X_{K-1} + \frac{1}{2} X_{K-1} X_K \mp \frac{i}{2} Y_{K-1} \pm \frac{i}{2} Y_{K-1} X_K \, ,
\end{equation}
Applying the shift operators leads to the reduced iso-spectral Hamiltonian in the $(K - 1)$ qubit space
\begin{equation}
1 \otimes H_{N}^{(K-1)} = S_+^{(K) \dagger} H^{(K)} S_+^{(K)} + S_-^{(K) \dagger} H^{(K)} S_-^{(K)} \, .
\label{eq:shift_operator}
\end{equation}
The above operation can be applied iteratively until the number of qubits cannot be reduced further. We will demonstrate our scheme in the following two examples.

\section{The Fermi--Hubbard Model}

As a first example, we examine the two-site Fermi--Hubbard model. Here, the Hamiltonian in second quantization is given by
\begin{eqnarray}
H_{\rm HM}^{(4)} &=& -t \left( c^\dagger_{1} c^{\phantom\dagger}_{2} + c^\dagger_{2} c^{\phantom\dagger}_{1} + c^\dagger_{3} c^{\phantom\dagger}_{4} + c^\dagger_{4} c^{\phantom\dagger}_{3} \right) + U \left( c^\dagger_{1} c^{\phantom\dagger}_{1} c^\dagger_{4} c^{\phantom\dagger}_{4} + c^\dagger_{2} c^{\phantom\dagger}_{2} c^\dagger_{3} c^{\phantom\dagger}_{3} \right) \, .
\label{eq:
_Hubbard_start}
\end{eqnarray}
This model consists of four states (different subscripts of the creation and annihilation operators) that can be assigned in the following way: qubit 1 corresponds to spin-up on the first site, qubit 2 to spin-up on the second site, qubit 3 to spin-down on the first site, and qubit 4 to spin-down on the second site. Using this mapping, applying the Jordan--Wigner transformation and exploiting the properties of the Pauli matrices, we obtain the Hamiltonian
\begin{eqnarray} \label{eq:HMJW}
H_{\rm HM}^{(4)} &=& - \frac{1}{2} t \left( X_1 X_2 + X_3 X_4 + Y_1 Y_2 + Y_3 Y_4 \right) \\
&+& \frac{1}{4} U \left( 2 + Z_1 Z_4 + Z_2 Z_3 - Z_1 - Z_2 - Z_3 - Z_4 \right) \nonumber \, .
\end{eqnarray}
For illustration, we write the matrix representation of this Hamiltonian 
\begin{equation}
\hspace{-2.5cm} H_{\rm HM}^{(4)} = \left(
\begin{array}{cccccccccccccccc}
 0 & 0 & 0 & 0 & 0 & 0 & 0 & 0 & 0 & 0 & 0 & 0 & 0 & 0 & 0 & 0 \\
 0 & 0 & -t & 0 & 0 & 0 & 0 & 0 & 0 & 0 & 0 & 0 & 0 & 0 & 0 & 0 \\
 0 & -t & 0 & 0 & 0 & 0 & 0 & 0 & 0 & 0 & 0 & 0 & 0 & 0 & 0 & 0 \\
 0 & 0 & 0 & 0 & 0 & 0 & 0 & 0 & 0 & 0 & 0 & 0 & 0 & 0 & 0 & 0 \\
 0 & 0 & 0 & 0 & 0 & 0 & 0 & 0 & -t & 0 & 0 & 0 & 0 & 0 & 0 & 0 \\
 0 & 0 & 0 & 0 & 0 & 0 & -t & 0 & 0 & -t & 0 & 0 & 0 & 0 & 0 & 0 \\
 0 & 0 & 0 & 0 & 0 & -t & U & 0 & 0 & 0 & -t & 0 & 0 & 0 & 0 & 0 \\
 0 & 0 & 0 & 0 & 0 & 0 & 0 & U & 0 & 0 & 0 & -t & 0 & 0 & 0 & 0 \\
 0 & 0 & 0 & 0 & -t & 0 & 0 & 0 & 0 & 0 & 0 & 0 & 0 & 0 & 0 & 0 \\
 0 & 0 & 0 & 0 & 0 & -t & 0 & 0 & 0 & U & -t & 0 & 0 & 0 & 0 & 0 \\
 0 & 0 & 0 & 0 & 0 & 0 & -t & 0 & 0 & -t & 0 & 0 & 0 & 0 & 0 & 0 \\
 0 & 0 & 0 & 0 & 0 & 0 & 0 & -t & 0 & 0 & 0 & U & 0 & 0 & 0 & 0 \\
 0 & 0 & 0 & 0 & 0 & 0 & 0 & 0 & 0 & 0 & 0 & 0 & 0 & 0 & 0 & 0 \\
 0 & 0 & 0 & 0 & 0 & 0 & 0 & 0 & 0 & 0 & 0 & 0 & 0 & U & -t & 0 \\
 0 & 0 & 0 & 0 & 0 & 0 & 0 & 0 & 0 & 0 & 0 & 0 & 0 & -t & U & 0 \\
 0 & 0 & 0 & 0 & 0 & 0 & 0 & 0 & 0 & 0 & 0 & 0 & 0 & 0 & 0 & 2 U \\
\end{array}
\right) \, ,
\end{equation}
which has dimension $16 \times 16$. Here, the Hamiltonian still contains all blocks corresponding to different numbers of electrons. For example, the element 0 in the top left corner belongs to 0 electrons and the element $2 U$ in the bottom right corner belongs to 4 electrons. 

Only states with two electrons are non-trivial and of physical interest for us. Therefore, in the next step, we project out all irrelevant states with the number of electrons different from two by using the projector
\begin{equation}
P_2^{(4)} = -\frac{1}{4} N_{\rm tot} (1 - N_{\rm tot}) (3 - N_{\rm tot}) (4 - N_{\rm tot}) \, .
\end{equation}
This leads to the new Hamiltonian:
\begin{equation}
H_{{\rm HM},2}^{(4)} = P_2^{(4)\dagger} H_{\rm HM}^{(4)} P_2^{(4)} \, .
\end{equation}
With the above-mentioned mapping between states (electrons) and qubits together with the Jordan--Wigner transformation, the projector can be written in terms of the Pauli operators:
\begin{eqnarray}
P_2^{(4)} &=& \frac{3}{8} - \frac{1}{8} Z_1 Z_2 - \frac{1}{8} Z_1 Z_3 - \frac{1}{8} Z_2 Z_3 - \frac{1}{8} Z_1 Z_4 \nonumber \\ 
&-& \frac{1}{8} Z_2 Z_4 - \frac{1}{8} Z_3 Z_4 + \frac{3}{8} Z_1 Z_2 Z_3 Z_4 \, .
\end{eqnarray}
Applying the projectors $P_2^{(4)}$ to Eq.~(\ref{eq:HMJW}) we obtain the following Hamiltonian, which formally operates exclusively on the two-electron subspace 
\begin{eqnarray}
H_{{\rm HM},2}^{(4)} &=& -\frac{t}{4} \left( X_1 X_2 + X_3 X_4 + Y_1 Y_2 + Y_3 Y_4 - Z_1 Z_2 X_3 X_4  \right. \nonumber \\ 
&-& \left. Z_1 Z_2 Y_3 Y_4 - X_1 X_2 Z_3 Z_4 - Y_1 Y_2 Z_3 Z_4 \right) \nonumber \\
&+& \frac{U}{8} \left( 1 - Z_1 Z_2 - Z_1 Z_3 + Z_2 Z_3 + Z_1 Z_4 - Z_2 Z_4 \right. \nonumber \\
&-& \left. Z_3 Z_4 + Z_1 Z_2 Z_3 Z_4 \right) \, .
\label{eq:h24}
\end{eqnarray}
Concerning the number of terms to compute, we have an increase from 11 to 16. However, moving to the matrix representation, we notice that only a $6 \times 6$  block along the diagonal has nonzero entries:
\begin{equation}
\hspace{-1.5cm} H_{{\rm HM},2}^{(4)} = \left(
\begin{array}{cccccccccccccccc}
 0 & 0 & 0 & 0 & 0 & 0 & 0 & 0 & 0 & 0 & 0 & 0 & 0 & 0 & 0 & 0 \\
 0 & 0 & 0 & 0 & 0 & 0 & 0 & 0 & 0 & 0 & 0 & 0 & 0 & 0 & 0 & 0 \\
 0 & 0 & 0 & 0 & 0 & 0 & 0 & 0 & 0 & 0 & 0 & 0 & 0 & 0 & 0 & 0 \\
 0 & 0 & 0 & 0 & 0 & 0 & 0 & 0 & 0 & 0 & 0 & 0 & 0 & 0 & 0 & 0 \\
 0 & 0 & 0 & 0 & 0 & 0 & 0 & 0 & 0 & 0 & 0 & 0 & 0 & 0 & 0 & 0 \\
 0 & 0 & 0 & 0 & 0 & 0 & -t & 0 & 0 & -t & 0 & 0 & 0 & 0 & 0 & 0 \\
 0 & 0 & 0 & 0 & 0 & -t & U & 0 & 0 & 0 & -t & 0 & 0 & 0 & 0 & 0 \\
 0 & 0 & 0 & 0 & 0 & 0 & 0 & 0 & 0 & 0 & 0 & 0 & 0 & 0 & 0 & 0 \\
 0 & 0 & 0 & 0 & 0 & 0 & 0 & 0 & 0 & 0 & 0 & 0 & 0 & 0 & 0 & 0 \\
 0 & 0 & 0 & 0 & 0 & -t & 0 & 0 & 0 & U & -t & 0 & 0 & 0 & 0 & 0 \\
 0 & 0 & 0 & 0 & 0 & 0 & -t & 0 & 0 & -t & 0 & 0 & 0 & 0 & 0 & 0 \\
 0 & 0 & 0 & 0 & 0 & 0 & 0 & 0 & 0 & 0 & 0 & 0 & 0 & 0 & 0 & 0 \\
 0 & 0 & 0 & 0 & 0 & 0 & 0 & 0 & 0 & 0 & 0 & 0 & 0 & 0 & 0 & 0 \\
 0 & 0 & 0 & 0 & 0 & 0 & 0 & 0 & 0 & 0 & 0 & 0 & 0 & 0 & 0 & 0 \\
 0 & 0 & 0 & 0 & 0 & 0 & 0 & 0 & 0 & 0 & 0 & 0 & 0 & 0 & 0 & 0 \\
 0 & 0 & 0 & 0 & 0 & 0 & 0 & 0 & 0 & 0 & 0 & 0 & 0 & 0 & 0 & 0 \\
\end{array}
\right) \, .
\end{equation}
The next step consists in shrinking the Hilbert space and eliminating the sector on which the operator $H_{{\rm HM},2}^{(4)}$ acts trivially as a zero operator. This operation corresponds to the removal of a qubit and the consequent reduction of the size of the Hilbert space from $2^4$ to $2^3$, and is obtained through the action of the shift operators
\begin{equation}
S_\pm^{(4)} = \frac{1}{2} X_3 \pm \frac{1}{2} X_3 X_4 \mp \frac{i}{2} Y_3 \pm \frac{i}{2} Y_3 X_4 \, .
\end{equation}
The name {\em shift operator} arises from the effect these operators have on the matrix representation of the Hamiltonian, $H_{{\rm HM},2}^{(4)}$: $S_+^{(4)}$ shifts a copy of the inner $8 \times 8$ block to the upper left corner of the matrix, whereas $S_{-}^{(4)}$ does the same but to the lower right. Applying the shift operators to $H_{{\rm HM},2}^{(4)}$, we get
\begin{equation}
1 \otimes H_{{\rm HM},2}^{(3)}  = S_+^{(4) \dagger} H_{{\rm HM},2}^{(4)} S_+^{(4)} + S_-^{(4) \dagger} H_{{\rm HM},2}^{(4)} S_-^{(4)} \, ,
\end{equation}
which simplifies to
\begin{eqnarray}
H_{{\rm HM},2}^{(3)} &=& -\frac{t}{2} \left( X_1 X_2 +Y_1 Y_2 - Z_1 Z_2 X_3 + X_3 \right) \nonumber \\
&+& \frac{U}{4} \left( 1 - Z_1 Z_2 +  Z_1 Z_3 -  Z_2 Z_3 \right) \, 
\end{eqnarray}
or, in matrix representation,
\begin{equation}
H_{{\rm HM},2}^{(3)} = \left(
\begin{array}{cccccccc}
 0 & 0 & 0 & 0 & 0 & 0 & 0 & 0 \\
 0 & 0 & -t & 0 & 0 & -t & 0 & 0 \\
 0 & -t & U & 0 & 0 & 0 & -t & 0 \\
 0 & 0 & 0 & 0 & 0 & 0 & 0 & 0 \\
 0 & 0 & 0 & 0 & 0 & 0 & 0 & 0 \\
 0 & -t & 0 & 0 & 0 & U & -t & 0 \\
 0 & 0 & -t & 0 & 0 & -t & 0 & 0 \\
 0 & 0 & 0 & 0 & 0 & 0 & 0 & 0 \\
\end{array}
\right) \, .
\end{equation}
The inner block of this $8 \times 8$ matrix still has the dimension $6 \times 6$. To reduce the number of qubits further (going from dimension $2^3$ to $2^2$), we need to reorder the states and create a non-zero block of dimension $4 \times 4$. The reordering (which in the matrix representation corresponds to a swap of the second column and row with the fourth) is achieved through the action of the reorder operator,
\begin{equation}
R^{(3)} = \frac{1}{2} \left( 1 +  Z_1 Z_3 -  Z_1 X_2 Z_3 + X_2 \right) \, .
\end{equation}
Note that this reorder operator is for this specific system and that for systems with higher dimensionality and a different number of electrons, a generalization has to be found. The new reordered three-qubit Hamiltonian is
\begin{equation}
H_{{\rm HM},2,{\rm R}}^{(3)} = R^{(3) \dagger} H_{{\rm HM},2}^{(3)} R^{(3)} \, ,
\end{equation}
which, after some algebra, becomes
\begin{eqnarray}
H_{{\rm HM},2,{\rm R}}^{(3)} &=& -\frac{t}{2} \left( X_2 X_3 + Y_2 Y_3 - X_1 Z_2 Z_3 + X_1 \right)\nonumber \\ 
&+& \frac{U}{4} \left( 1 - Z_1 Z_2 + Z_1 Z_3 - Z_2 Z_3 \right)
\end{eqnarray}
and in matrix form
\begin{equation}
H_{{\rm HM},2,{\rm R}}^{(3)} =
\left(
\begin{array}{cccccccc}
 0 & 0 & 0 & 0 & 0 & 0 & 0 & 0 \\
 0 & 0 & 0 & 0 & 0 & 0 & 0 & 0 \\
 0 & 0 & U & -t & -t & 0 & 0 & 0 \\
 0 & 0 & -t & 0 & 0 & -t & 0 & 0 \\
 0 & 0 & -t & 0 & 0 & -t & 0 & 0 \\
 0 & 0 & 0 & -t & -t & U & 0 & 0 \\
 0 & 0 & 0 & 0 & 0 & 0 & 0 & 0 \\
 0 & 0 & 0 & 0 & 0 & 0 & 0 & 0 \\
\end{array}
\right) \, .
\end{equation}
As in the previous dimensional reduction step, we reduce the dimension of the relevant Hilbert space by applying the pair of shift operators,
\begin{equation}
S_\pm^{(3)} = \frac{1}{2} X_2 \pm \frac{1}{2} X_2 X_3 \mp \frac{i}{2} Y_2 \pm \frac{i}{2} Y_2 X_3 \, ,
\end{equation}
to the Hamiltonian $H_{{\rm HM},2,{\rm R}}^{(3)} $. This leads to
\begin{equation}
1 \otimes H_{{\rm HM},2}^{(2)} = S_+^{(3) \dagger} H_{{\rm HM},2,{\rm R}}^{(3)} S_+^{(3)} + S_-^{(3) \dagger} H_{{\rm HM},2,{\rm R}}^{(3)} S_-^{(3)}
\end{equation}
and finally to the two-qubit Hamiltonian 
\begin{equation}
H_{{\rm HM},2}^{(2)} =  -t \left( X_1 + X_2 \right) + \frac{U}{2} \left( 1 +  Z_1 Z_2 \right) \, .
\end{equation}
In matrix representation this is
\begin{equation}
H_{{\rm HM},2}^{(2)} = 
\left(
\begin{array}{cccc}
 U & -t & -t & 0 \\
 -t & 0 & 0 & -t \\
 -t & 0 & 0 & -t \\
 0 & -t & -t & U \\
\end{array}
\right) \, .
\end{equation}
The final reduced Hamiltonian is iso-spectral to the Hamilton in Eq.~(\ref{eq:h24}) and has the following eigenvalues
\begin{equation}
E_{1,2,3,4} = 0, U, \frac{1}{2} \left(U \pm \sqrt{16 t^2 + U^2}\right) \, .
\end{equation}
On a quantum computer this Hamiltonian is very simple to simulate as there is only a $Z_1 Z_2$ interaction between the qubits. The many-body interactions of order 4 of the original Hamiltonian are therefore trivially reduced to order of 2. These interactions can easily be trotterized or even adiabatically simulated on the quantum computer with the correct tunable couplings~\cite{biamonte_realizable_2008}. We have therefore shown how reduce the Hamiltonian of the Fermi--Hubbard model from four to three to two qubits without having to go into the basis.

\section{The Hydrogen Molecule} 

Here, we apply  the same dimensional reduction scheme to the hydrogen molecule. The Hamiltonian of the hydrogen molecule with two single-electron molecular orbitals and after the Jordan--Wigner transformation is
\begin{eqnarray}
H_{H_2}^{(4)} &=& f_1 + f_2 Z_1 Z_2 + f_2 Z_3 Z_4 + f_3 Z_1 Z_3 \nonumber\\
&+& f_3 Z_2 Z_4 + f_4 Z_2 Z_3 + f_5 Z_1 Z_4 + f_6 X_1 X_2 X_3 X_4 \nonumber\\
&+& f_6 X_1 X_2 Y_3 Y_4 + f_6 Y_1 Y_2 X_3 X_4 + f_6 Y_1 Y_2 Y_3 Y_4 \nonumber\\
&+& f_7 Z_1 + f_7 Z_4 + f_8 Z_2 + f_8 Z_3 \, .
\end{eqnarray}
The eight coefficients are simple functions of the one-body and two-body (Coulomb) orbital integrals $h_{ij}$ and $h_{ijkl}$, which define the Hamiltonian in  Eq.~(\ref{eq:Hamiltonian}),
\begin{eqnarray}
f_1 &=& h_{11}-\frac{h_{1212}}{2}+h_{1221}+\frac{h_{1441}}{4}+h_{22}+\frac{h_{2332}}{4} \,, \nonumber \\
f_2 &=& \frac{h_{1221}}{4}-\frac{h_{1212}}{4} \,,\quad
f_3 = \frac{h_{1221}}{4} \,, \nonumber \\
f_4 &=& \frac{h_{2332}}{4} \,, \quad
f_5 = \frac{h_{1441}}{4} \,,\quad
f_6 = \frac{h_{1212}}{4} \,, \nonumber \\
f_7 &=& -\frac{h_{11}}{2}+\frac{h_{1212}}{4}-\frac{h_{1221}}{2}-\frac{h_{1441}}{4} \,, \nonumber \\
f_8 &=& \frac{h_{1212}}{4}-\frac{h_{1221}}{2}-\frac{h_{22}}{2}-\frac{h_{2332}}{4} \,
\end{eqnarray}
and can be precomputed using a classical computer~\cite{whitfield_simulation_2011}. To be consistent with the Fermi--Hubbard model, the qubits are mapped in the same way as above: qubit 1 corresponds to a spin-up electron in the first molecular orbital, qubit 2 to a spin-up electron in the second molecular orbital, qubit 3 to a spin-down electron in the first molecular orbital, and finally qubit 4 to a spin-down electron in the second molecular orbital. Note that this mapping differs from the one reported in Refs.~\cite{aspuru-guzik_simulated_2005, whitfield_simulation_2011}.

We now apply the same sequence of operations as used in the Fermi--Hubbard case to reduce the number of qubits required for the simulation of the hydrogen molecule. In addition to the reduction of the Hilbert space to a two-electron sector, we will also restrict the system to a spin singlet. The projector then is given by
\begin{equation}
P_2^{(4)} = N_\uparrow (2 - N_\uparrow) N_\downarrow (2 - N_\downarrow) \, ,
\end{equation}
where
\begin{equation}
N_\uparrow = n_1 + n_2 \quad {\rm and}  \quad N_\downarrow = n_3 + n_4 \, .
\end{equation}
The projector in terms of the Pauli operators is
\begin{equation}
P_2^{(4)} = \frac{1}{4} - \frac{1}{4} Z_1 Z_2 - \frac{1}{4} Z_3 Z_4 + \frac{1}{4} Z_1 Z_2 Z_3 Z_4 \, .
\end{equation}
After applying the projector $P_2^{(4)}$, the Hamiltonian is reduced to three qubits first. 
The resulting Hamiltonian is then reordered and finally reduced to a $2^2$ dimensional Hilbert space through the action of the reduction operation, which leads to
\begin{eqnarray}
H_{H_2}^{(2)} &=& f_1 - 2 f_2 + 4 f_6 X_1 X_2 + ( -2 f_3 + f_4 + f_5 ) Z_1 Z_2 \nonumber \\
&+& (f_7 - f_8) Z_1 + (f_7 - f_8) Z_2 \, .
\label{eq:h22q}
\end{eqnarray}
In matrix representation this becomes (Appendix \ref{appendix:c})
\begin{equation}
H_{H_2}^{(2)} = 
\left(
\begin{array}{cccc}
{\small \begin{array}{c} f_1 - 2 f_2 \hfill \\ - 2 f_3 + f_4 \hfill \\ + f_5 + 2 f_7 \hfill \\ -2 f_8 \hfill \end{array}} & 0 & 0 & 4 f_6 \\
 0 & {\small \begin{array}{c} f_1 - 2 f_2 \hfill \\ + 2 f_3 - f_4 \hfill \\ - f_5 \hfill \end{array}} & 4 f_6 & 0 \\
 0 & 4 f_6 & {\small \begin{array}{c} f_1 - 2 f_2 \hfill \\ + 2 f_3 - f_4 \hfill \\ - f_5 \hfill \end{array}} & 0 \\
 4 f_6 & 0 & 0 & {\small  \begin{array}{c} f_1 - 2 f_2 \hfill \\ - 2 f_3 + f_4 \hfill \\ + f_5 - 2 f_7 \hfill \\ + 2 f_8 \hfill \end{array}} \\
\end{array}
\right) \, .
\end{equation}
Using the parameter from Ref.~\cite{whitfield_simulation_2011} for the overlap integrals, we obtain the well-known eigenvalues for the hydrogen molecule for this minimal basis set
\begin{eqnarray}
E_{1,2,3,4} &=& -1.85105 \, {\rm Ha}, -1.24623 \, {\rm Ha}, -0.88365 \, {\rm Ha}, -0.23389 \, {\rm Ha} \, .
\end{eqnarray}
In contrast to the Fermi--Hubbard model, the Hamiltonian of the hydrogen molecule in Eq.~(\ref{eq:h22q}) a $X_1 X_2$ interaction is needed in addition to the $Z_1 Z_2$ interaction between the two qubits. Still, this Hamiltonian could be simulated on a current quantum computer with moderate effort.

When reducing from three to two qubits, a reordering has to be performed for both Hamiltonians. This reordering depends on the way the quantum states are mapped onto the qubits. For larger systems, the blocks of the same number of electrons are distributed in the Hamiltonian in a complicated, but, regular fashion. Therefore, with the appropriate book-keeping, it should be possible to generalize the reduction of qubits for larger quantum systems. However, for most Hamiltonians this reordering will exponentially hard and scale with $2^K$.

\section{Conclusion}

In conclusion, we have described a scheme to reduce the number of qubits required for the simulation of a fermionic system described by a Hamiltonian with a Hilbert space of $2^K$. We exploit the fact that in second quantization the corresponding Hilbert space is given by the direct sum of the subspaces, each corresponding to a fixed number of electrons. Importantly, this scheme can be carried out in operator space. For the Fermi--Hubbard model and the hydrogen molecule, we introduced a scheme to dimension of the Hilbert space to a subspace characterized by a fixed number of electrons. The is achieved at the cost of an increase of the number of terms in the Hamiltonian, which in the worst case scales with $2^K$. However, the dimensionality of Hilbert space will be reduced from $2^{K}$ to $K!/((K - N)! N!)$. In both examples shown, the number of qubits was reduced from four to two. Our scheme involves no approximations, and the reduction comes only from identifying the relevant parts of the Hilbert space together with the proper procedure to carry out the reduction. As a result, the two physical systems can successfully be simulated on a quantum computer based on just two qubits with moderate effort. The next step would be to apply the reduction scheme to molecules with a moderate number of 10 to 20 qubits and study the growth of the number of terms in Hamiltonian. Furthermore, for these small system size it should be examined if the reorder operator can be generalized. Even with both mentioned limitations, the proposed scheme still remains a useful tool to reduce the dimensionality of specific quantum systems for quantum simulators with a limited number of resources.

\ack

We thank S.~Filipp for valuable comments on the manuscript.

\appendix

\section{Block diagonality of the Hamiltonian}

The total number of electrons for each element of a four-state Hamiltonian that corresponds to the cross sum of the state is:
\begin{equation}
\left(
\begin{array}{cccccccccccccccc}
 0 & \circ & \circ & \circ & \circ & \circ & \circ & \circ & \circ & \circ & \circ & \circ & \circ & \circ & \circ & \circ \\
 \circ & 1 & 1 & \circ & 1 & \circ & \circ & \circ & 1 & \circ & \circ & \circ & \circ & \circ & \circ & \circ \\
 \circ & 1 & 1 & \circ & 1 & \circ & \circ & \circ & 1 & \circ & \circ & \circ & \circ & \circ & \circ & \circ \\
 \circ & \circ & \circ & 2 & \circ & 2 & 2 & \circ & \circ & 2 & 2 & \circ & 2 & \circ & \circ & \circ \\
 \circ & 1 & 1 & \circ & 1 & \circ & \circ & \circ & 1 & \circ & \circ & \circ & \circ & \circ & \circ & \circ \\
 \circ & \circ & \circ & 2 & \circ & 2 & 2 & \circ & \circ & 2 & 2 & \circ & 2 & \circ & \circ & \circ \\
 \circ & \circ & \circ & 2 & \circ & 2 & 2 & \circ & \circ & 2 & 2 & \circ & 2 & \circ & \circ & \circ \\
 \circ & \circ & \circ & \circ & \circ & \circ & \circ & 3 & \circ & \circ & \circ & 3 & \circ & 3 & 3 & \circ \\
 \circ & 1 & 1 & \circ & 1 & \circ & \circ & \circ & 1 & \circ & \circ & \circ & \circ & \circ & \circ & \circ \\
 \circ & \circ & \circ & 2 & \circ & 2 & 2 & \circ & \circ & 2 & 2 & \circ & 2 & \circ & \circ & \circ \\
 \circ & \circ & \circ & 2 & \circ & 2 & 2 & \circ & \circ & 2 & 2 & \circ & 2 & \circ & \circ & \circ \\
 \circ & \circ & \circ & \circ & \circ & \circ & \circ & 3 & \circ & \circ & \circ & 3 & \circ & 3 & 3 & \circ \\
 \circ & \circ & \circ & 2 & \circ & 2 & 2 & \circ & \circ & 2 & 2 & \circ & 2 & \circ & \circ & \circ \\
 \circ & \circ & \circ & \circ & \circ & \circ & \circ & 3 & \circ & \circ & \circ & 3 & \circ & 3 & 3 & \circ \\
 \circ & \circ & \circ & \circ & \circ & \circ & \circ & 3 & \circ & \circ & \circ & 3 & \circ & 3 & 3 & \circ \\
 \circ & \circ & \circ & \circ & \circ & \circ & \circ & \circ & \circ & \circ & \circ & \circ & \circ & \circ & \circ & 4 \\
\end{array}
\right)
\end{equation}
No electrons are created or annihilated. Only if the number of electrons of both states is the same does the matrix element exists. The other matrix elements are denoted by the symbol $\circ$. 

The elements after projection, so that only elements with two electrons and a singlet configuration survive
\begin{equation}
\left(
\begin{array}{cccccccccccccccc}
 \circ & \circ & \circ & \circ & \circ & \circ & \circ & \circ & \circ & \circ & \circ & \circ & \circ & \circ & \circ & \circ \\
 \circ & \circ & \circ & \circ & \circ & \circ & \circ & \circ & \circ & \circ & \circ & \circ & \circ & \circ & \circ & \circ \\
 \circ & \circ & \circ & \circ & \circ & \circ & \circ & \circ & \circ & \circ & \circ & \circ & \circ & \circ & \circ & \circ \\
 \circ & \circ & \circ & \circ & \circ & \circ & \circ & \circ & \circ & \circ & \circ & \circ & \circ & \circ & \circ & \circ \\
 \circ & \circ & \circ & \circ & \circ & \circ & \circ & \circ & \circ & \circ & \circ & \circ & \circ & \circ & \circ & \circ \\
 \circ & \circ & \circ & \circ & \circ & 2 & 2 & \circ & \circ & 2 & 2 & \circ & \circ & \circ & \circ & \circ \\
 \circ & \circ & \circ & \circ & \circ & 2 & 2 & \circ & \circ & 2 & 2 & \circ & \circ & \circ & \circ & \circ \\
 \circ & \circ & \circ & \circ & \circ & \circ & \circ & \circ & \circ & \circ & \circ & \circ & \circ & \circ & \circ & \circ \\
 \circ & \circ & \circ & \circ & \circ & \circ & \circ & \circ & \circ & \circ & \circ & \circ & \circ & \circ & \circ & \circ \\
 \circ & \circ & \circ & \circ & \circ & 2 & 2 & \circ & \circ & 2 & 2 & \circ & \circ & \circ & \circ & \circ \\
 \circ & \circ & \circ & \circ & \circ & 2 & 2 & \circ & \circ & 2 & 2 & \circ & \circ & \circ & \circ & \circ \\
 \circ & \circ & \circ & \circ & \circ & \circ & \circ & \circ & \circ & \circ & \circ & \circ & \circ & \circ & \circ & \circ \\
 \circ & \circ & \circ & \circ & \circ & \circ & \circ & \circ & \circ & \circ & \circ & \circ & \circ & \circ & \circ & \circ \\
 \circ & \circ & \circ & \circ & \circ & \circ & \circ & \circ & \circ & \circ & \circ & \circ & \circ & \circ & \circ & \circ \\
 \circ & \circ & \circ & \circ & \circ & \circ & \circ & \circ & \circ & \circ & \circ & \circ & \circ & \circ & \circ & \circ \\
 \circ & \circ & \circ & \circ & \circ & \circ & \circ & \circ & \circ & \circ & \circ & \circ & \circ & \circ & \circ & \circ \\
\end{array}
\right) \, .
\end{equation}

\section{Tensor operations to reduce the number of qubits}
\label{appendix:b}

A generic one-qubit Hamiltonian in matrix form is
\begin{equation}
H^{(1)} = \left(
\begin{array}{cc}
H_{11} & H_{12}\\
H_{21} & H_{22}\\
\end{array}
\right) \, .
\end{equation}
Enlarging it by one qubit to two qubits gives
\begin{eqnarray}
H^{(2)} &=& 1 \otimes H^{(1)} \\
&=& 1 \otimes \left(
\begin{array}{cc}
H_{11} & H_{12}\\
H_{21} & H_{22}\\
\end{array}
\right)  \nonumber \\
&=& \left(
\begin{array}{cccc}
H_{11} & H_{12} & 0 & 0\\
H_{21} & H_{22} & 0 & 0\\
0 & 0 & H_{11} & H_{12}\\
0 & 0 & H_{21} & H_{22}\\
\end{array}
\right) \nonumber \\
&=& H^{(1)} \oplus H^{(1)} \, .
\end{eqnarray}

In our scheme, we want to do the opposite: We want to reduce the Hamiltonian by one qubit. The starting point is a two-qubit Hamiltonian $H_{\rm ib}^{(2)}$ with a $4 \times 4$ inner block surrounded by zeros:
\begin{equation}
H_{\rm ib}^{(2)} = \left(
\begin{array}{cccc}
0 & 0 & 0 & 0\\
0 & H_{11} & H_{12} & 0\\
0 & H_{21} & H_{22} & 0\\
0 & 0 & 0 & 0\\
\end{array}
\right) \, .
\end{equation}
This Hamiltonian can be reduced by the following operation, which shifts a copy of the inner block up and a copy down:
\begin{eqnarray}
1 \otimes H^{(1)}
&=& S_+^{(2) \dagger} H_{\rm ib}^{(2)} S_+^{(2)} + S_-^{(2) \dagger} H_{\rm ib}^{(2)} S_-^{(2)} \, .
\label{eq:block}
\end{eqnarray}
The two-qubit operator, which shifts the inner block up and down, is given by
\begin{equation}
S_\pm^{(2)} = \frac{1}{2} X_1 \pm \frac{1}{2} X_1 X_2 \mp \frac{i}{2} Y_1 \pm \frac{i}{2} Y_1 X_2 \, .
\end{equation}
In matrix form, the shift-up operator is
\begin{equation}
S_+^{(2)} = \left(
\begin{array}{cccc}
 0 & 0 & 0 & 1 \\
 1 & 0 & 0 & 0 \\
 0 & 1 & 0 & 0 \\
 0 & 0 & 1 & 0 \\
\end{array}
\right) \, ,
\end{equation}
and the shift-down operator is
\begin{equation}
S_-^{(2)} = \left(
\begin{array}{cccc}
 0 & 1 & 0 & 0 \\
 0 & 0 & 1 & 0 \\
 0 & 0 & 0 & 1 \\
 1 & 0 & 0 & 0 \\
\end{array}
\right) \, .
\end{equation}
Inserting the shift operators into Eq.~(\ref{eq:block}) gives
\begin{eqnarray}
1 \otimes H_{\rm ib}^{(1)} &=& \left(
\begin{array}{cccc}
H_{11} & H_{12} & 0 & 0\\
H_{21} & H_{22} & 0 & 0\\
0 & 0 & 0 & 0\\
0 & 0 & 0 & 0\\
\end{array}
\right) + \left(
\begin{array}{cccc}
0 & 0 & 0 & 0\\
0 & 0 & 0 & 0\\
0 & 0 & H_{11} & H_{12}\\
0 & 0 & H_{21} & H_{22}\\
\end{array}
\right) \nonumber \\
&=& \left(
\begin{array}{cccc}
H_{11} & H_{12} & 0 & 0\\
H_{21} & H_{22} & 0 & 0\\
0 & 0 & H_{11} & H_{12}\\
0 & 0 & H_{21} & H_{22}\\
\end{array}
\right) \nonumber \\
&=& 1 \otimes \left(
\begin{array}{cc}
H_{11} & H_{12}\\
H_{21} & H_{22}\\
\end{array}
\right) \, .
\end{eqnarray}
Finally, the two-qubit Hamiltonian $H_{\rm ib}^{(2)}$ is reduced to a one-qubit Hamiltonian $H_{\rm ib}^{(1)}$.

\newpage

\section{The intermediate steps of qubit reduction for the hydrogen molecule}
\label{appendix:c}

Below, all intermediate equations for the reduction of qubits for the hydrogen molecule are shown, which have been omitted. The four-qubit Hamiltonian of the hydrogen molecule in matrix form is
\begin{equation}
\hspace{-3.5cm} H_{H_2}^{(4)} =
\arraycolsep=-2pt
{\footnotesize \left(
\begin{array}{cccccccccccccccc}
 {\scriptsize \begin{array}{c}\, \,f_1+2 f_2\hfill\\+2 f_3+f_4\hfill\\+f_5+2 f_7\hfill\\+2 f_8\hfill\end{array}} & 0 & 0 & 0 & 0 & 0 & 0 & 0 & 0 & 0 & 0 & 0 & 0 & 0 & 0 & 0 \\
 0 & {\scriptsize \begin{array}{c}f_1+f_4\hfill\\-f_5+2 f_8\hfill\end{array}} & 0 & 0 & 0 & 0 & 0 & 0 & 0 & 0 & 0 & 0 & 0 & 0 & 0 & 0 \\
 0 & 0 & {\scriptsize \begin{array}{c}f_1-f_4\hfill\\+f_5+2 f_7\hfill\end{array}} & 0 & 0 & 0 & 0 & 0 & 0 & 0 & 0 & 0 & 0 & 0 & 0 & 0 \\
 0 & 0 & 0 & {\scriptsize \begin{array}{c}f_1+2 f_2\hfill\\-2 f_3-f_4\hfill\\-f_5\hfill\end{array}} & 0 & 0 & 0 & 0 & 0 & 0 & 0 & 0 & 0 & 0 & 0 & 0 \\
 0 & 0 & 0 & 0 & {\scriptsize \begin{array}{c}f_1-f_4\hfill\\+f_5+2 f_7\hfill\end{array}} & 0 & 0 & 0 & 0 & 0 & 0 & 0 & 0 & 0 & 0 & 0 \\
 0 & 0 & 0 & 0 & 0 & {\scriptsize \begin{array}{c}f_1-2 f_2\hfill\\+2 f_3-f_4\hfill\\-f_5\hfill\end{array}} & 0 & 0 & 0 & 0 & 4 f_6 & 0 & 0 & 0 & 0 & 0 \\
 0 & 0 & 0 & 0 & 0 & 0 & {\scriptsize \begin{array}{c}f_1-2 f_2\hfill\\-2 f_3+f_4\hfill\\+f_5+2 f_7\hfill\\-2 f_8\hfill\end{array}} & 0 & 0 & 4 f_6 & 0 & 0 & 0 & 0 & 0 & 0 \\
 0 & 0 & 0 & 0 & 0 & 0 & 0 & {\scriptsize \begin{array}{c}f_1+f_4\hfill\\-f_5-2 f_8\hfill\end{array}} & 0 & 0 & 0 & 0 & 0 & 0 & 0 & 0 \\
 0 & 0 & 0 & 0 & 0 & 0 & 0 & 0 & {\scriptsize \begin{array}{c}f_1+f_4\hfill\\-f_5+2 f_8\hfill\end{array}} & 0 & 0 & 0 & 0 & 0 & 0 & 0 \\
 0 & 0 & 0 & 0 & 0 & 0 & 4 f_6 & 0 & 0 & {\scriptsize \begin{array}{c}f_1-2 f_2\hfill\\-2 f_3+f_4\hfill\\+f_5-2 f_7\hfill\\+2 f_8\hfill\end{array}} & 0 & 0 & 0 & 0 & 0 & 0 \\
 0 & 0 & 0 & 0 & 0 & 4 f_6 & 0 & 0 & 0 & 0 & {\scriptsize \begin{array}{c}f_1-2 f_2\hfill\\+2 f_3-f_4\hfill\\-f_5\hfill\end{array}} & 0 & 0 & 0 & 0 & 0 \\
 0 & 0 & 0 & 0 & 0 & 0 & 0 & 0 & 0 & 0 & 0 & {\scriptsize \begin{array}{c}f_1-f_4\hfill\\+f_5-2 f_7\hfill\end{array}} & 0 & 0 & 0 & 0 \\
 0 & 0 & 0 & 0 & 0 & 0 & 0 & 0 & 0 & 0 & 0 & 0 & {\scriptsize \begin{array}{c}f_1+2 f_2\hfill\\-2 f_3-f_4\hfill\\-f_5\hfill\end{array}} & 0 & 0 & 0 \\
 0 & 0 & 0 & 0 & 0 & 0 & 0 & 0 & 0 & 0 & 0 & 0 & 0 & {\scriptsize \begin{array}{c}f_1-f_4\hfill\\+f_5-2 f_7\hfill\end{array}} & 0 & 0 \\
 0 & 0 & 0 & 0 & 0 & 0 & 0 & 0 & 0 & 0 & 0 & 0 & 0 & 0 & {\scriptsize \begin{array}{c}f_1+f_4\hfill\\-f_5-2 f_8\hfill\end{array}} & 0 \\
 0 & 0 & 0 & 0 & 0 & 0 & 0 & 0 & 0 & 0 & 0 & 0 & 0 & 0 & 0 & {\scriptsize \begin{array}{c}f_1+2 f_2\hfill\\+2 f_3+f_4\hfill\\+f_5-2 f_7\hfill\\-2 f_8\hfill\end{array}} \\
\end{array}
\right) \, . }
\end{equation}
When applying the projector $P_2^{(4)}$, which restricts the space to two electrons, we obtain following Hamiltonian:
\begin{eqnarray}
H_{H_2,2}^{(4)} &=& \frac{f_1}{4} - \frac{f_2}{2} + \left( -\frac{f_1}{4} + \frac{f_2}{2} \right) Z_1 Z_2 + \left( \frac{f_3}{2} - \frac{f_4}{4} - \frac{f_5}{4} \right) Z_1 Z_3 \nonumber \\
&+& \left( -\frac{f_3}{2} + \frac{f_4}{4} + \frac{f_5}{4} \right) Z_2 Z_3 + \left( -\frac{f_3}{2} + \frac{f_4}{4} + \frac{f_5}{4} \right) Z_1 Z_4  \nonumber \\
&+& \left( \frac{f_3}{2} - \frac{f_4}{4} - \frac{f_5}{4} \right) Z_2 Z_4 - \frac{1}{4} f_1 Z_3 Z_4 + \frac{1}{2} f_2 Z_3 Z_4  \nonumber \\ 
&+& \left( \frac{f_7}{4} - \frac{f_8}{4} \right) Z_1 Z_2 Z_3 + \left( -\frac{f_7}{4} + \frac{f_8}{4} \right) Z_1 Z_2 Z_4  \nonumber \\ 
&+& \left( -\frac{f_7}{4} + \frac{f_8}{4} \right) Z_1 Z_3 Z_4 + \left( \frac{f_7}{4} - \frac{f_8}{4} \right) Z_2 Z_3 Z_4  \nonumber \\
&+& f_6 X_1 X_2 X_3 X_4 + f_6 Y_1 Y_2 X_3 X_4 + f_6 X_1 X_2 Y_3 Y_4 + f_6 Y_1 Y_2 Y_3 Y_4  \nonumber \\
&+& \left( \frac{f_1}{4} - \frac{f_2}{2} \right) Z_1 Z_2 Z_3 Z_4 + \left( \frac{f_7}{4} - \frac{f_8}{4} \right) Z_1 + \left( -\frac{f_7}{4} + \frac{f_8}{4} \right) Z_2  \nonumber \\
&+& \left( -\frac{f_7}{4} + \frac{f_8}{4} \right) Z_3 + \left( \frac{f_7}{4} - \frac{f_8}{4} \right) Z_4 \, .
\end{eqnarray}
In matrix form, the Hamiltonian is
\begin{equation}
\hspace{-2.5cm} H_{H_2,2}^{(4)} =
\arraycolsep=3pt
\left(
\begin{array}{cccccccccccccccc}
 0 & 0 & 0 & 0 & 0 & 0 & 0 & 0 & 0 & 0 & 0 & 0 & 0 & 0 & 0 & 0 \\
 0 & 0 & 0 & 0 & 0 & 0 & 0 & 0 & 0 & 0 & 0 & 0 & 0 & 0 & 0 & 0 \\
 0 & 0 & 0 & 0 & 0 & 0 & 0 & 0 & 0 & 0 & 0 & 0 & 0 & 0 & 0 & 0 \\
 0 & 0 & 0 & 0 & 0 & 0 & 0 & 0 & 0 & 0 & 0 & 0 & 0 & 0 & 0 & 0 \\
 0 & 0 & 0 & 0 & 0 & 0 & 0 & 0 & 0 & 0 & 0 & 0 & 0 & 0 & 0 & 0 \\
 0 & 0 & 0 & 0 & 0 & {\scriptsize \begin{array}{c}f_1-2 f_2\hfill\\+2 f_3-f_4\hfill\\-f_5\hfill\end{array}} & 0 & 0 & 0 & 0 & 4 f_6 & 0 & 0 & 0 & 0 & 0 \\
 0 & 0 & 0 & 0 & 0 & 0 & {\scriptsize \begin{array}{c}f_1-2 f_2\hfill\\-2 f_3+f_4\hfill\\+f_5+2 f_7\hfill\\-2 f_8\hfill\end{array}} & 0 & 0 & 4 f_6 & 0 & 0 & 0 & 0 & 0 & 0 \\
 0 & 0 & 0 & 0 & 0 & 0 & 0 & 0 & 0 & 0 & 0 & 0 & 0 & 0 & 0 & 0 \\
 0 & 0 & 0 & 0 & 0 & 0 & 0 & 0 & 0 & 0 & 0 & 0 & 0 & 0 & 0 & 0 \\
 0 & 0 & 0 & 0 & 0 & 0 & 4 f_6 & 0 & 0 & {\scriptsize \begin{array}{c}f_1-2 f_2\hfill\\-2 f_3+f_4\hfill\\+f_5-2 f_7\hfill\\+2 f_8\hfill\end{array}} & 0 & 0 & 0 & 0 & 0 & 0 \\
 0 & 0 & 0 & 0 & 0 & 4 f_6 & 0 & 0 & 0 & 0 & {\scriptsize \begin{array}{c}f_1-2 f_2\hfill\\+2 f_3-f_4\hfill\\-f_5\hfill\end{array}} & 0 & 0 & 0 & 0 & 0 \\
 0 & 0 & 0 & 0 & 0 & 0 & 0 & 0 & 0 & 0 & 0 & 0 & 0 & 0 & 0 & 0 \\
 0 & 0 & 0 & 0 & 0 & 0 & 0 & 0 & 0 & 0 & 0 & 0 & 0 & 0 & 0 & 0 \\
 0 & 0 & 0 & 0 & 0 & 0 & 0 & 0 & 0 & 0 & 0 & 0 & 0 & 0 & 0 & 0 \\
 0 & 0 & 0 & 0 & 0 & 0 & 0 & 0 & 0 & 0 & 0 & 0 & 0 & 0 & 0 & 0 \\
 0 & 0 & 0 & 0 & 0 & 0 & 0 & 0 & 0 & 0 & 0 & 0 & 0 & 0 & 0 & 0 \\
\end{array}
\right) \, .
\end{equation}
This Hamiltonian, which only exhibits a inner block of a $6 \times 6$ matrix, can be reduced to three qubits by applying the shift operators:
\begin{eqnarray}
H_{H_2,2}^{(3)} &=& \frac{f_1}{2} - f_2 + \left( -\frac{f_1}{2} + f_2 \right) Z_1 Z_2 + \left( - f_3 + \frac{f_4}{2} + \frac{f_5}{2} \right) Z_1 Z_3  \nonumber \\
&+& \left( f_3 - \frac{f_4}{2} - \frac{f_5}{2} \right) Z_2 Z_3 + 2 f_6 X_1 X_2 X_3 + 2 f_6 Y_1 Y_2 X_3 \nonumber \\
&+& \left( -\frac{f_7}{2} + \frac{f_8}{2} \right) Z_1 Z_2 Z_3 + \left( \frac{f_7}{2} - \frac{f_8}{2} \right) Z_1 \nonumber \\
&+& \left( -\frac{f_7}{2} + \frac{f_8}{2} \right) Z_2 + \left( \frac{f_7}{2} - \frac{f_8}{2} \right) Z_3 \, .
\end{eqnarray}
This Hamiltonian corresponds to a $8 \times 8$ matrix
\begin{equation}
\hspace{-2.5cm} H_{H_2,2}^{(3)} =
\left(
\begin{array}{cccccccc}
 0 & 0 & 0 & 0 & 0 & 0 & 0  \\
 0 & {\scriptsize \begin{array}{c}f_1-2 f_2\hfill\\+2 f_3-f_4\hfill\\-f_5\hfill\end{array}} & 0 & 0 & 0 & 0 & 4 f_6 & 0  \\
 0 & 0 & {\scriptsize \begin{array}{c}f_1-2 f_2\hfill\\-2 f_3+f_4\hfill\\+f_5+2 f_7\hfill\\-2 f_8\hfill\end{array}} & 0 & 0 & 4 f_6 & 0 & 0  \\
 0 & 0 & 0 & 0 & 0 & 0 & 0 & 0  \\
 0 & 0 & 0 & 0 & 0 & 0 & 0 & 0  \\
 0 & 0 & 4 f_6 & 0 & 0 & {\scriptsize \begin{array}{c}f_1-2 f_2\hfill\\-2 f_3+f_4\hfill\\+f_5-2 f_7\hfill\\+2 f_8\hfill\end{array}} & 0 & 0  \\
 0 & 4 f_6 & 0 & 0 & 0 & 0 & {\scriptsize \begin{array}{c}f_1-2 f_2\hfill\\+2 f_3-f_4\hfill\\-f_5\hfill\end{array}} & 0  \\
 0 & 0 & 0 & 0 & 0 & 0 & 0 & 0  \\
\end{array}
\right) \, .
\end{equation}
As in the case of the Fermi--Hubbard model, this Hamiltonian has to be reordered to be reduced in size further. The reordered Hamiltonian is
\begin{eqnarray}
H_{H_2,2,{\rm R}}^{(3)} &=& \frac{f_1}{2} - f_2 + \left( f_3 - \frac{f_4}{2} - \frac{f_5}{2} \right) Z_1 Z_2 + \left( - f_3 + \frac{f_4}{2} + \frac{f_5}{2} \right) Z_1 Z_3 \nonumber \\
&+& \left( -\frac{f_1}{2} + f_2 \right) Z_2 Z_3 + 2 f_6 X_1 X_2 X_3 + 2 f_6 X_1 Y_2 Y_3 \nonumber \\
&+& \left( -\frac{f_7}{2} + \frac{f_8}{2} \right) Z_1 Z_2 Z_3 + \left( \frac{f_7}{2} - \frac{f_8}{2} \right) Z_1 \nonumber \\ 
&+& \left( -\frac{f_7}{2} + \frac{f_8}{2} \right) Z_2 + \left( \frac{f_7}{2} - \frac{f_8}{2} \right) Z_3
\end{eqnarray}
and in matrix form
\begin{equation}
\hspace{-2.5cm} H_{H_2,2,{\rm R}}^{(3)} =
\left(
\begin{array}{cccccccc}
 0 & 0 & 0 & 0 & 0 & 0 & 0 & 0  \\
 0 & 0 & 0 & 0 & 0 & 0 & 0 & 0  \\
 0 & 0 & {\scriptsize \begin{array}{c} f_1 - 2 f_2 \\ - 2 f_3 + f_4 \hfill \\ + f_5 + 2 f_7 \hfill \\ -2 f_8 \hfill \end{array}} & 0 & 0 & 4 f_6 & 0 & 0 \\
 0 & 0 & 0 & {\scriptsize \begin{array}{c} f_1 - 2 f_2\\ + 2 f_3 - f_4 \hfill \\ - f_5 \hfill \end{array}} & 4 f_6 & 0 & 0 & 0 \\
 0 & 0 & 0 & 4 f_6 & {\scriptsize \begin{array}{c} f_1 - 2 f_2 \\ + 2 f_3 - f_4 \hfill \\ - f_5 \hfill \end{array}} & 0 & 0 & 0 \\
 0 & 0 & 4 f_6 & 0 & 0 & {\scriptsize \begin{array}{c} f_1 - 2 f_2 \\ - 2 f_3 + f_4 \hfill \\ + f_5 - 2 f_7 \hfill \\ + 2 f_8 \hfill \end{array}} & 0 & 0 \\
 0 & 0 & 0 & 0 & 0 & 0 & 0 & 0  \\
 0 & 0 & 0 & 0 & 0 & 0 & 0 & 0  \\
\end{array}
\right) \, .
\end{equation}
This Hamiltonian can now be reduced to two qubits by applying again the shift operators. The two-qubit Hamiltonian of the hydrogen molecule is iso-spectral to the original Hamiltonian. It has the following eigenvalues:
\begin{eqnarray}
E_{1,2,3,4} &=& f_1 - 2 f_2 + 2 f_3 - f_4 - f_5 - 4 f_6 \, , \nonumber \\
&& f_1 - 2 f_2 + 2 f_3 - f_4 - f_5 + 4 f_6 \, , \nonumber \\
&& f_1 - 2 f_2 - 2 f_3 + f_4 + f_5 - 2 \sqrt{4 f_6^2 + f_7^2 - 2 f_7 f_8 + f_8^2} \, , \nonumber \\
&& f_1 - 2 f_2 - 2 f_3 + f_4 + f_5 + 2 \sqrt{4 f_6^2 + f_7^2 - 2 f_7 f_8 + f_8^2} \, .
\end{eqnarray}

\section*{References}

\bibliographystyle{iopart-num}

\bibliography{qubits}

\end{document}